\documentclass[conference]{IEEEtran}
\IEEEoverridecommandlockouts

\usepackage[
    sorting=none
]{biblatex}

\usepackage{amsmath,amssymb,amsfonts}
\usepackage{algorithm}
\usepackage[noend]{algpseudocode}
\usepackage{algcompatible}

\makeatletter
\def\BState{\State\hskip-\ALG@thistlm}
\makeatother
\usepackage{graphicx}
\usepackage{textcomp}
\usepackage{xcolor}
\usepackage{mathtools}

\usepackage{tikz}
\usetikzlibrary{shapes.geometric, arrows}
\usepackage{float}
\usepackage[utf8]{inputenc} 
\usepackage[T1]{fontenc} 
\usepackage{fouriernc} 
\usepackage{multirow}
\usepackage{relsize}

\def\BibTeX{{\rm B\kern-.05em{\sc i\kern-.025em b}\kern-.08em
    T\kern-.1667em\lower.7ex\hbox{E}\kern-.125emX}}

\usepackage[printwatermark]{xwatermark}
\usepackage{xcolor}
\usepackage{graphicx}
\usepackage{lipsum}

\newcommand{\fref}[1]{Fig.~\ref{#1}}

\newcommand{\eref}[1]{Eq.~(\ref{#1})}

\begin{document}
\title{Non-contact Infant Sleep Apnea Detection\\
}

\author{
\IEEEauthorblockN{Gihan Jayatilaka*, Harshana Weligampola*, Suren Sritharan*, Pankayraj Pathmanathan,\\ Roshan Ragel, Isuru Nawinne}
 \IEEEauthorblockA
 {Department of Computer Engineering \\
 Faculty of Engineering, University of Peradeniya, Sri Lanka\\
\{ gihanjayatilaka,harshana.w,suren.sri \} @ eng.pdn.ac.lk , \{e14237,roshanr,isurun \}  @ce.pdn.ac.lk  
}}

\maketitle

\begin{abstract}
Sleep apnea is a breathing disorder where a person repeatedly stops breathing in sleep. Early detection is crucial for infants because it might bring long term adversities. The existing accurate detection mechanism (pulse oximetry) is a skin contact measurement. The existing non-contact mechanisms (acoustics, video processing) are not accurate enough. This paper presents a novel algorithm for the detection of sleep apnea with video processing. The solution is non-contact, accurate and lightweight enough to run on a single board computer. The paper discusses the accuracy of the algorithm on real data, advantages of the new algorithm, its limitations and suggests future improvements.
\end{abstract}
\footnote{*Equally contributing authors}


\section{Introduction}
\subsection{Sleep Apnea}

Sleep Apnea~\cite{sleep-apnea,sleep-apnea-review} is a sleeping disorder caused by the interruption of breathing during sleep.
An affected individual may stop breathing for a short time. Repetitive occurrence will result in low blood oxygen content which can lead to other complications. It can affect people of any age, but infants with this condition it can have serious adverse effects. If untreated it could have long term adverse affects on the infant. Thus early detection is crucial.

There are two forms of Sleep Apnea -- obstructive sleep apnea (OSA) and central sleep apnea. Obstructive sleep apnea occurs because of a physical disturbance to the nasal passage. Central sleep apnea occurs because of a neurological disorder.

\subsection{Risk groups}
The general risk groups~\cite{sleep-apnea} of sleep apnea condition are over weight people, males with long necks, people with abnormalities in the neck. Down Syndrome has a correlation with sleep apnea in both adults and children. Children with large tonsils are also a risk group~\cite{sleep-apnea}.

Premature babies have a higher risk of sleep apnea and therefore this paper proposes a solution for that risk group.

\subsection{Effects}
Fluctuations in the blood oxygen levels, increased heart rates, elevated blood pressure and an increase of the risk of strokes are effects of sleep apnea. Mood changes and impaired concentration could be observed in sleep apnea victims.

When sleep apnea occurs in an infant~\cite{sleep-apnea-infants}, these effects (mainly the dip in blood oxygen level) can harm the brain growth causing long term effects.


\section{Available solutions}
\subsection{Pulse oximetry}
Pulse oximetry~\cite{pulse-oxy} is a technique of measuring the oxygen concentration of the blood. If a child is suffering from OSA his/her blood oxygen level drops suddenly. Sometimes the drop in the blood oxygen level could be a result of some other problem as well. But still, since the biggest problem with OSA is the reduction of the oxygen supply to the brain, pulse oximeters are good enough as an OSA detection technique.

Pulse oximeters should be connected to the skin with underlying veins present. Usually, they are connected to the fingers in adults. But for infants, they are connected to the earlobes. 

\subsection{Acoustics}
Acoustic techniques~\cite{acoustics} used to detect OSA consists of microphones and sound processing. The microphones tries to sense the sound of the breathing of the infant. These sound signals are then processed to identify anomalies in the pattern.

The disadvantage of this technique is that the sound of breathing is of low strength and therefore the noise in the background can make it very difficult to analyze the breathing sound.

\subsection{Video Processing}
Feasibility of using video analysis to detect sleep apnea has been proved through manual analysis~\cite{video-manual}. Later works attempt to automate this analysis.\\
\begin{itemize}
    \item \textbf{Individual pixel time series analysis}~\cite{mexico}\\
    The approach is to find the rate of breathing by the Fourier transform of the time series of the grey scale values of a set of pixels. This algorithm is sufficient for cases where the only movement in the video is breathing. External disturbances will result in anomalous readings when using this technique.
    
    \item \textbf{Regional average colour intensity time series analysis}~\cite{iit}\\
    The algorithm is based on keeping track of the total \textbf{intensity} of a region of interest. The problem with this algorithm is that it does not take into account any actual shape of the infant. The intensity of the region can change drastically with lighting condition changes. These changes are taken as false positives in the algorithm.
    
\end{itemize}

These approaches do not give accurate results because they do not take into account the fact that breathing pattern of a baby is clearly seen through the movement of some parts of the body. 

\section{Proposed solution}
The solution consists of 3 steps : (1) Identifying the infant, (2) measure breathing pattern, (3) detect anomalies in breathing pattern. Each step is described in detail in this section.

The infant identification is done by training a \textbf{custom built} neural network. Breathing pattern measurement utilized a \textbf{modified} version of canny edge detection algorithm for robustness and a \textbf{novel algorithm} for finding the breathing cycles. The anomalies are detected by a threshold for longer time intervals between breaths.

\subsection{Infant detection}

The breathing pattern detection algorithm is applied on a localized region. This region of interest is identified using a convolution neural network. This information is used to rotate the camera using the servo motors as intended, in order to cover the whole region of interest (stomach region), after which the breathing pattern could be analyzed.

For our case, we need to only identify the probability of existence of an infant in the frame ($P_{c}$) and the localization information (width ($b_{w}$)  and height ($b_{h}$) of the infant in the picture in pixels and the coordinate of the center of the infant ($b_{x}, b_{y}$)). Therefore, we defined the target label $y$ as in \eref{eq:neural-out}.

\begin{equation}
\label{eq:neural-out}
y = \left[ 
    \begin{tabular}{c}
        $P_{c}$\\
        $b_{x}$\\
        $b_{y}$\\
        $b_{w}$\\
        $b_{h}$\\
    \end{tabular}
    \right ]
\end{equation}

We created a data-set that contains images of infants with age below 6 months and labeled each image with previously defined parameters. Also, some random images that do not contain any infant were added and labeled with $P_{c} = 0$. These images with infants and without infants were distributed with a ratio of 3:1 in the data-set.

\begin{figure}[H]
    \centering
    \includegraphics[scale=0.6]{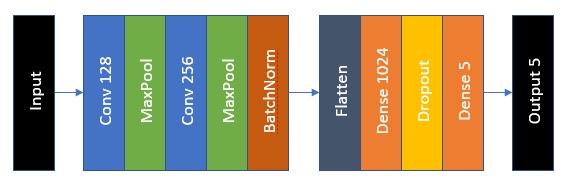}
    \caption{Neural network model}
    \label{fig:neural-network}
\end{figure}

The neural network \(\fref{fig:neural-network}\) consist of convolution and max pooling layers to activate on interesting regions and reduce the complexity. A deeper network is not necessary because we need to identify only a single class. Then the output feature matrix of those layers was flattened. Dense layers are used to predict the target label $y$.

The dataset was divided into validation set and training set with 1:4 ratio. Training data were used to train the model with batch size of 64 for 1000 epochs.

After training, this model can be used to localize the infant in a given image.

\subsection{Breathing pattern analysis}

\subsubsection{Edge detection}
    
    The 3 colour video input of 8 bit depth is converted into the greyscale video $G(t)_{[H\times W]}$. The grey-scale video is fed to a modified version of \textbf{Canny edge detection algorithm}~\cite{canny,opencv_library}
    
    The first step of the algorithm is to blur the image using a Gaussian kernel of size $5 \times 5$ to filter the noise that can give false positives as edges. The kernel $K_{blur}$ is used to convolute as per \eref{eq:kblur}.
    
    \begin{equation}
    \label{eq:kblur}
    \begin{aligned}
    K_{blur} &= \frac{1}{159}
    \begin{bmatrix} 
    2&4&5&4&2 \\
    4&9&12&9&4 \\
    5&12&15&12&5 \\
    4&9&12&9&4 \\
    2&4&5&4&2
    \end{bmatrix}\\
    G_1(t) &= G(t) \circledast K_{blur}
    \end{aligned}
    \end{equation}
    
    Then the derivatives of this grey-scale intensity matrix is obtained along $x$ and $y$ axes by convolution with $K_{diff x}$ and $K_{diff y}$ kernels as in \eref{eq:grad-conv}.
    
    \begin{equation}
    \label{eq:grad-conv}
    \begin{aligned}
    K_{diff x} &= 
    \begin{bmatrix}
         -1&0&1 \\
         -2&0&2 \\
         -1&0&1
    \end{bmatrix}\\
    K_{diff y} &= 
    \begin{bmatrix}
         1&2&1  \\
         0&0&0\\
         -1&-2&-1
    \end{bmatrix}\\
    D_x(t) &= G_1(t) \circledast K_{diff x}\\
    D_y(t) &= G_1(t) \circledast K_{diff y}
    \end{aligned}
    \end{equation}
    
    By considering these two matrices to contain the vector components of the gradients, the resultant gradient vectors for $(t,x,y)$ is calculated as the matrice $D(t)$ which contains the magnitudes of the particular vectors and $\theta(t)$ which contains the angles of the particular vectors are calculated by \eref{eq:grad-mat}.

    
    \begin{equation}
    \begin{aligned}
    \label{eq:grad-mat}
    D(t,x,y)&=\sqrt{D_x(t,x,y)^2 + D_y(t,x,y)^2}\\ 
    \theta(t,x,y)&=tan^{-1} \frac{D_y(t,x,y)}{D_x(t,x,y)} \end{aligned}        
    \end{equation}

    The gradients are discretized into 4 directions as $$\left \{ \left [0,\frac{\pi}{8} \right ) \cup \left [\frac{7 \pi}{8},\pi \right ) , \left [\frac{\pi}{8},\frac{3 \pi}{8} \right ) , \left [\frac{3 \pi}{8},\frac{5 \pi}{8} \right ) , \left [\frac{5 \pi}{8},\frac{7 \pi}{8} \right) \right \}$$
    The $\theta(t)$ matrix is replaced by $\left \{0 , \frac{\pi}{4}, \frac{\pi}{2} , \frac{3\pi}{4} \right \}$ for convenience.\\
    Non-maximum suppression~\cite{non-max} is applied on these directions to find thin edges from these gradient matrices.

    Let the thin edges obtained be ${E_{thin}(t)}_{H\times W}$. This matrix undergoes a threshold step to generate the final edge matrix ${E_{(t)}}_{H\times W}$. Thresholding step requires two threshold values for $D(t,x,y)$ as $T_{high}(t)$ and $T_{low}(t)$. These are calculated by using the mean $(\mu_{(t)})$ and the standard deviation $(\sigma_{(t)})$ of the thin edge matrix as per equations given in \eref{eq:ttt}.
    
    
    
    \begin{IEEEeqnarray}{rl}\label{eq:ttt}
    \mu_{(t)} & = \frac{1}{H \times W} \sum _{x=1,y=1} ^{x=W,y=H} E_{thin}(t,x,y)   \IEEEyesnumber \IEEEyessubnumber\\
    \sigma_{(t)} & = \sqrt{\frac{1}{H\times W} \sum _{x=1,y=1} ^{x=W,y=H} (E_{thin}(t,x,y) - \mu_{(t)})^2} \IEEEyessubnumber\\
    T_{high}(t)& = \mu_{(t)}+0.5\sigma_{(t)} \IEEEyessubnumber\\
    T_{low}(t)& = \mu_{(t)}-0.5\sigma_{(t)}\IEEEyessubnumber
    \end{IEEEeqnarray}

\begin{algorithm}
\caption{Edge threshold algorithm}\label{}
\begin{algorithmic}[1]
\FOR{$x \in [1,W]$}
\FOR{$y \in [1,H]$}
    \IF{$E_{thin}(t,x,y) < T_{low}(t)$}
    $E(t,x,y)=0$
    \ELSIF{$E_{thin}(t,x,y) > T_{high}(t)$}
    $E(t,x,y)=1$
    \ELSE
    $E(t,x,y)= NA$
    \ENDIF
\ENDFOR
\ENDFOR

\FOR{$x \in [1,W]$}
\FOR{$y \in [1,H]$}
    \IF{$T_{low}(t) < E_{thin}(t,x,y) < T_{high}(t)$}
    \IF{$\exists (x_0,y_0)$ such that $(x_0,y_0) \neq(x,y) , \|x-x_0\| \leq 1 , \|y-y_0\| \leq 1 , E(t,x_0,y_0)=1 $ }
    $E(t,x,y)=1$
    \ELSE
    $E(t,x,y)=0$
    \ENDIF
    \ENDIF
\ENDFOR
\ENDFOR

\end{algorithmic}
\end{algorithm}

The edge thresholding~\cite{opencv_library} is done by \textbf{Algorithm 1} which first recognizes the pixels with a gradient more than $T_{high}(t)$ as edges and the pixels with a gradient less than $T_{low}(t)$ to not be edges. Then the remaining pixels are categorized into edges only if they have a neighbouring edge pixel.   

Finally, the edge matrix $E_(t)_{H\times W}$ is obtained. This could be visualized as \fref{fig:region-interest}.
    
    $$E_{(t,x,y)}=  \left \{ \begin{array}{l}
        1 \textrm{ if } E_{(t,x,y)} \textrm{ is an edge. }\\
        0 \textrm{ if } E_{(t,x,y)} \textrm{ is not an edge. }\\ 
        \end{array} \right \} $$

\subsubsection{Choosing the area of interest}

\begin{figure}[H]
        \centering
        \frame{\includegraphics[keepaspectratio,scale=0.18]{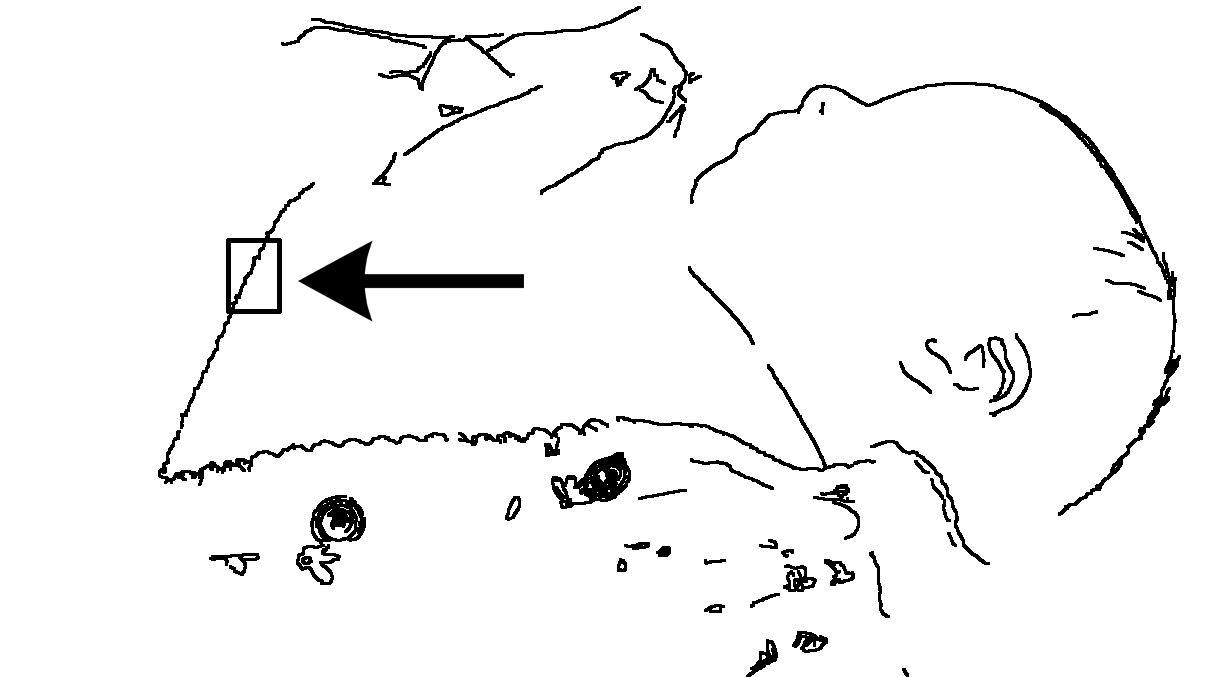}}
        \caption{The region of interest}
        \label{fig:region-interest}
    \end{figure}

    First the region of interest $A_0$ is chosen. This region is marked in \fref{fig:region-interest}.
    $$(x,y) \in A_0 $$
    $$x\in [ x_{0},x_{1} ] , y\in [ y_{0},y_{1} ]$$
    These values could be obtained by the relationships given in \eref{eq:xy} with the neural network output.
    \begin{IEEEeqnarray}{ll}\label{eq:xy}
    x_{0} = b_x - 0.5b_W \; \IEEEyesnumber \IEEEyessubnumber\\
    x_{1} = b_x + 0.5b_W \IEEEyessubnumber \\
    y_{0} = b_y - 0.5b_H \IEEEyessubnumber \\
    y_{1} = b_y + 0.5b_H \IEEEyessubnumber 
    \end{IEEEeqnarray}
    

\subsubsection{Center of gravity tracking}
    The centroid $C_{0}(t)=(x_{c_0(t)},y_{c_0(t)}))$ of the edges in $A_0$ is calculated for every $t$ by \eref{eq:cog}.

    \begin{IEEEeqnarray}{ll}\label{eq:cog}
    x_{c_0(t)} =\frac{ \sum _{(x,y) \in A} E(t,x,y) \times x}{ \sum _{(x,y) \in A} E(t,x,y)}\IEEEyesnumber \IEEEyessubnumber\\
    y_{c_0(t)} =\frac{ \sum _{(x,y) \in A} E(t,x,y) \times y}{ \sum _{(x,y) \in A} E(t,x,y)} \IEEEyessubnumber
   \end{IEEEeqnarray}
    $$\textrm{Special case : }
    (x_{c_0(t)},y_{c_0(t)}) = \left(\frac{x_0 + x_1}{2},\frac{y_0 + y_1}{2}\right) $$$$ \textrm{ when } \sum _{(x,y) \in A} E(t,x,y)  = 0$$

\subsubsection{Subspace filtering}
    Then the velocity of the centroid $\underline{v}_{(t)}$ is calculated by,\\
    $\underline{v}_{(t)} =(x_{c_0(t)}-x_{c_0(t-1)})\underline{i} + (y_{c_0(t)}-y_{c_0(t-1)})\underline{j} $
    
    The direction along which the velocity of the centroid  $\underline{v}_{(t)}$ lie is calculated using the \textbf{Principle component analysis} as follows,
    
    Write $v_{(t)}$ as a row vector calculated by \eref{eq:cog-vel}.
    \begin{equation}
    \begin{aligned}
    \label{eq:cog-vel}
    v_{(t)} &= \begin{pmatrix}
         \underline{v}_{(t)} . \underline{i} & \underline{v}_{(t)} . \underline{j}
    \end{pmatrix}\\
    v_{(t)} &= \begin{pmatrix}
         v_{x(t)} & v_{y(t)} 
    \end{pmatrix}
    \end{aligned}
    \end{equation}
    
    Make a matrix by taking 10 such readings and arranging them as rows in \eref{eq:10}.
    
    \begin{equation}
    \label{eq:10}
    V_{(t)} = 
    \begin{pmatrix}
        v_{x(t)} & v_{y(t)} \\
        v_{x(t-1)} & v_{y(t-1)} \\
        v_{x(t-2)} & v_{y(t-2)} \\....&...\\...&.....\\
        v_{x(t-9)} & v_{y(t-9)} \\
    \end{pmatrix}
    \end{equation}
    
    The row means are calculated by \eref{eq:means}.
    \begin{IEEEeqnarray}{ll}\label{eq:means}
    \overline{v_{x(t)}}=\frac{1}{10} \sum _{i=0}^{9},v_{x(t-i)}\IEEEyesnumber \IEEEyessubnumber \\ \overline{v_{y(t)}}=\frac{1}{10} \sum _{i=0}^{9} v_{y(t-i)} \IEEEyessubnumber \end{IEEEeqnarray}
    
    Then the difference matrix $D_{(t)}$ and the co-variance matrix $C_{(t)}$ are calculated by \eref{eq:diff}, and \eref{eq:cov}.

    \begin{equation}
    \label{eq:diff}
    D_{(t)} = V_{(t)}-\overline{V_{(t)}} = 
    \begin{pmatrix}
         v_{x(t)}-\overline{v_{x(t)}} & v_{y(t)}-\overline{v_{y(t)}} \\
         v_{x(t-1)}-\overline{v_{x(t)}} & v_{y(t-1)}-\overline{v_{y(t)}} \\
         v_{x(t-2)}-\overline{v_{x(t)}} & v_{y(t-2)}-\overline{v_{y(t)}} \\....&...\\...&.....\\
         v_{x(t-9)}-\overline{v_{x(t)}} & v_{y(t-9)}-\overline{v_{y(t)}}
    \end{pmatrix}
    \end{equation}

    \begin{equation}
    \label{eq:cov}
    C_{(t)}=D_{(t)}^T .D_{(t)}
    \end{equation}

    $C_{(t)}$ is decomposed into $C_{(t)} = P_{(t)} D_{(t)} P_{(t)}^{-1}$ by the eigen value decomposition to give the matrices in \eref{eq:pdp}.
    
    \begin{equation}
    \begin{aligned}
    \label{eq:pdp}
    P_{(t)} &= 
    \begin{pmatrix}
         w_{1x(t)}&w_{2x(t)}  \\
         w_{1y(t)}& w_{2y(t)}
    \end{pmatrix}
    D_{(t)} &= 
    \begin{pmatrix}
         \lambda_{1(t)}&0  \\
         0& \lambda_{2(t)}
    \end{pmatrix}
    \end{aligned}
    \end{equation}
    
    Here the $P_{(t)}$ has the eigen vectors,\\
    $$\underline{w}_{1(t)}=w_{1x(t)}\underline{i}+w_{1y(t)}\underline{j}$$
    $$\underline{w}_{2(t)}=w_{2x(t)}\underline{i}+w_{2y(t)}\underline{j}$$
    
    $D_{(t)}$ has their corresponding eigen values $\lambda_{1(t)}$ and $\lambda_{2(t)}$.\\
    
    The bigger value of $\lambda_{1(t)}$ and $\lambda_{2(t)}$ is chosen (let it be $\lambda_{1(t)}$ ) and the corresponding eigen vector $\underline{w}_{1(t)}$ gives the direction of the breathing.
    
    The unit vector along this direction is calculated by dividing the vector by the second norm in \eref{eq:unit}.
    \begin{equation}
    \label{eq:unit}
    u{(t)}=\frac{\underline{w}_{1(t)}}{\|{\underline{w}_{1(t)}}\|}
    \end{equation}

\begin{figure}[H]
        \centering
        \includegraphics[width=8cm, height=3cm]{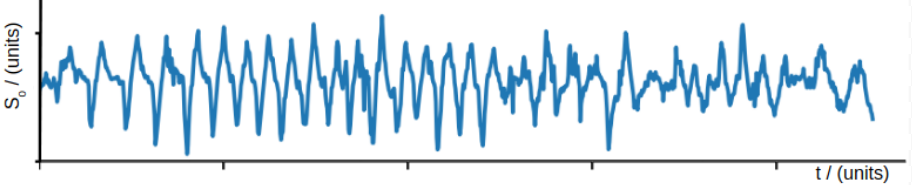}
        \caption{$s_{0(t)} \textrm{ versus } t$}
        \label{fig:s0}
    \end{figure}

    Now we have $\underline{u}{(t)}$ and $\underline{v}_{(t)}$. Projecting the velocity vector in the unit vector of direction as in \eref{eq:scalar} to give a scalar parameter $s_{0(t)}$ that can be used to determine breathing. But it has a coarse variation with time as in \fref{fig:s0}.
    
    \begin{equation}
    \label{eq:scalar}s_{0(t)}=\underline{u}{(t)} . \underline{v}_{(t)}\end{equation}
    
\subsubsection{Smoothing}

\begin{figure}[H]
        \centering
        \includegraphics[width=8cm, height=3cm]{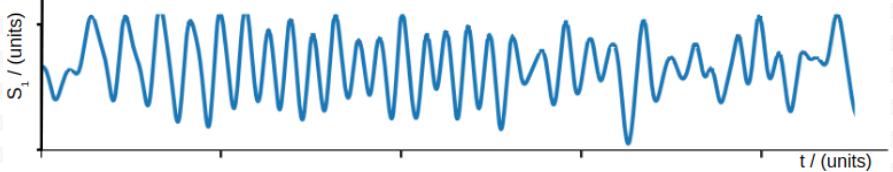}
        \caption{$s_{1(t)} \textrm{ versus } t$}
        \label{fig:s1}
    \end{figure}

    $s_{(0)t}$ undergoes two smoothing techniques in \eref{eq:smooth} to give $s_{1(t)}$ (\fref{fig:s1}) which is smoother and corresponds to the actual breathing pattern of the infant.
    \begin{equation}
    \begin{aligned}
    \label{eq:smooth}
    s_{1(t)} \leftarrow \textrm{ low pass filtered } s_{0(t)} \\
    s_{1(t)}=s_{0(t)}\times 0.8 + s_{0(t-1)}\times0.2 
    \end{aligned}
    \end{equation}
    
\subsubsection{Detecting the breathing intervals}
    The peaks are found using the technique,
    \begin{center} $s_{1(t)}>s_{1(t-1)} \textrm{ and } s_{1(t)} \leq s_{1(t+1)} \Rightarrow s_{1(t)} $ is a peak.\end{center}
    
\begin{figure}[H]
        \centering
        \includegraphics[width=8cm, height=3cm]{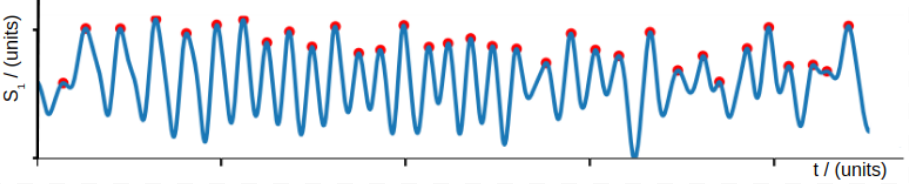}
        \caption{$s_{1(t)}$ with peaks marked}
        \label{fig:peaks}
    \end{figure}

    The identified peaks are marked in \fref{fig:peaks} and the time between them are considered to be breathing intervals as shown in Table~\ref{tab:breath-time}.
    
    \begin{table}[H]
            \caption{Breathing time interval report}

    \begin{center}\begin{tabular}{|c|c|}
        \hline
         Breath number& Time for cycle / (ms) \\
         \hline
         1&2200\\2&1830\\3&1950\\4&2050\\
         ...&...\\
         \hline
    \end{tabular}
        \label{tab:breath-time}
    \end{center}
    \end{table}

\subsection{Anomaly detection in the breathing pattern}
The infants can showcase periodic breathing pauses up to 10 seconds and it is not a health concern. A pause of more than 15 seconds would be a serious concern and could be categorized as sleep apnea~\cite{sleep-apnea}.

The proposed solution considers a window of 200 seconds (approximately 100 breaths). The algorithm rates the severity of condition as the number of high interval pauses. That is, the number of pauses more than 15 seconds during that interval. The severity value of each window is given as the output.

\section{Implementation of the solution}
\subsection{Implementation}
The full program and algorithms were implemented using Python programming language, open-CV library \cite{opencv_library}, sk-learn library \cite{scikit-learn}, numpy numerical computation package and matplotlib was used for all the output diagrams.

The hardware used for testing was raspberry pi single board computer. A raspberry pi camera working at 720p was used for video input and \textit{TowerPro SG90} servo motors were used to rotate this camera as per the output of the infant localization network.

\subsection{Performance}
The implemented system could work in real time. The algorithm is lightweight enough to run on a single board computer with 1.0 GHz processor and 1 GB RAM. The implementation does not use the GPU.

The system performed in real time for a 24 frames per second 720p video feed on the Raspberry pi 3 model B single board computer.

\section{Testing}
Testing was done in three steps (1) Infant detection, (2) Breathing pattern analysis and (3) Anomaly detection.\\
The video dataset was prepared by recording 8 infants' breathing patterns in the hospital. Each video was approximately 20 minutes. Four infants had sleep apnea and four did not. The video was manually marked for testing.\\ Another dataset (1000 images including 200 baby images) was created (by obtaining images from Google images and marking them) to train and test the infant detection algorithm. 

\subsection{Infant detection}

In order to test the algorithm, an area of interest (box) was manually marked for each frame. Then, the actual (manually marked) area of interest was compared with the area marked by the algorithm. If the actual area was within the marked area and the ratio of the areas was above 60\%, then the frame was classified as correctly marked (accuracy = 1), and if not it was incorrectly marked (accuracy = 0). The accuracy was calculated as the average accuracy of all the frames.

\subsection{Breathing pattern analysis}
The videos were analyzed and the high and the low points of the stomach were marked manually. The portion between a high and low point was considered to be a single segment. Thus two contiguous segments make up a complete breathing cycle. Then the peaks that were marked by the algorithm was superimposed with the breathing cycles. If there was exactly one peak per breathing cycle, then the output of the algorithm was considered to be correct for that cycle (accuracy = 1), otherwise the output was considered to be incorrect for that cycle (accuracy = 0).

\subsection{Anomaly detection}
The manual markings of data was compared with the algorithm results to categorize the algorithm results into four categories as true positive (TP), false positive (FP), true negative (TN) or false negative (FN). The detection accuracy ratio (DAR) and false alarm ratio (FAR) were calculated as in \eref{eq:dar}.

\begin{equation}
\label{eq:dar}
DAR =\frac{TP}{TP+TN}\hspace{0.3cm},\hspace{0.3cm}
FAR =\frac{FP}{FP+FN}
\end{equation}

\subsection{Percentile accuracy of tests}
\begin{table}[H]
\caption{Percentile accuracy of tests}

\begin{center}\begin{tabular}{|c|c|}
     \hline
     Test & Accuracy\\
     \hline
     Infant detection & 90\% \\
     Breathing pattern analysis& 86\%\\
     The detection accuracy ratio & 80\%\\
     False alarm ratio & 14\% \\
     \hline
\end{tabular}
        \label{tab:accuracy}
    \end{center}\end{table}   
        Table \ref{tab:accuracy} shows that the baby detection algorithm has a very high accuracy. But this accuracy requires good lighting conditions. Breathing pattern recognition (and therefore anomaly detection) is of comparable but less accuracy because sometimes the baby's movement is not happening only because of breathing.
        
\section{Discussion}
     Proposed solution is non-intrusive and non-contact. This makes it convenient for the patient as well as the medical staff. The algorithm is automated than the previous work found as the previous algorithms require human intervention to detect interesting regions.
     
    The algorithm is lightweight and can run on a cheap single board computer. The cost of the system is very low.
    
    The \textit{modified canny edge detection algorithm} is robust than the standard canny edge detection algorithm because of adaptive thresholding. The new \textit{breathing detection algorithm} is accurate than the existing algorithms.

\section{Future work}
    The algorithm performance could be increased by implementing it in parallel so that the single board computers can process 1080p video input in real time. This will allow more accurate and precise results.
    
    Current system requires periodic update of \textit{area of interest} by the neural network to provide error free output when the infant moves/turns. A reactive procedure could be implemented in place of this by an intelligent algorithm.
    
    The accuracy of the breathing pattern recognition could be improved by by fusing multiple areas of interest. The accuracy of the anomaly detection algorithm could be improved by incorporating a dataset consisting the breathing patterns of healthy infants and infants suffering from sleep apnea.
    
    The proposed solution should be tested with medical supervision, benchmarked against previous work assessed for standards before deployment.
    
\section*{Acknowledgment}

The authors acknowledge the contribution by Anupamali Willamuna (Department of Electrical and Electronic Engineering, University of Peradeniya) and Nuwan Jaliyagoda (Deaprtment of Computer Engineering, University of Peradeniya) in designing an embedded system that run the proposed algorithm and testing it. A part of the dataset was collected by Mr. Dinidu Bhathiya (Sysco Labs).

\printbibliography

\vspace{12pt}

\end{document}